\documentclass[aps,showpacs,twocolumn,floatfix]{revtex4} 
\usepackage{graphicx}
\begin{document}
   
%\twocolumn[\hsize\textwidth\columnwidth\hsize\csname@twocolumnfalse\endcsname
 
\title{Stabilization of a    light bullet in a layered 
Kerr medium with  sign-changing
nonlinearity}

%\title{Stable  three-dimensional soliton by 
%Kerr-nonlinearity management}

\author{Sadhan K. Adhikari}
\affiliation{Instituto de F\'{\i}sica Te\'orica, Universidade Estadual
Paulista, 01.405-900 S\~ao Paulo, S\~ao Paulo, Brazil}
%\address{Instituto de F\'{\i}sica Te\'orica, Universidade Estadual
%Paulista,  01.405-900 S\~ao Paulo, S\~ao Paulo, Brazil}

\date{\today}
%\maketitle

\begin{abstract}

Using the numerical solution of the nonlinear Schr\"odinger equation and a
variational method it is shown that (3+1)-dimensional spatiotemporal
optical solitons, known as light bullets, can be stabilized in a layered
Kerr medium with sign-changing nonlinearity along the propagation
direction.

\end{abstract}
\pacs{42.65.Jx, 42.65.Tg}
\maketitle

\section{Introduction}

After the  prediction of self-trapping
\cite{st} of an
optical beam in a nonlinear medium resulting in an optical soliton
\cite{0,1}, there
have been many
theoretical and experimental studies  to stabilize such
a soliton under different conditions of nonlinearity.   A bright soliton
in (1+1) dimension (D)  in Kerr medium is unconditionally stable for
positive or
self-focusing (SF)
nonlinearity in the nonlinear Schr\"odinger equation
(NLS) \cite{1}. 
However, in
(2+1)D in homogeneous 
bulk Kerr medium one cannot have a stable soliton-like axisymmetric 
cylindrical beam
\cite{2,3,3a}.
Also, in (3+1)D in such a  medium one cannot have a stable
optical wave
packet that remain confined in all   directions. Such a
confined wave packet in (3+1)D is often called a light bullet
and represents the extension of a self-trapped optical beam into the
temporal domain \cite{1}. 
If the nonlinearity is negative or self-defocusing (SDF), any initially
created
soliton spreads out in both (2+1)D and (3+1)D \cite{1}. If the
nonlinearity is
positive or of SF type, any initially created soliton is unstable and
eventually collapses \cite{1}.

Recently,  through a numerical simulation as well as a  variational
calculation 
based on the NLS  it has
been shown 
that
the axisymmetric
cylindrical beam in (2+1)D  
can be stabilized in a layered medium if a variable
nonlinearity coefficient is used in different layers \cite{mal,ber}. A
weak
modulation of
the nonlinearity coefficient along the propagation direction leads to a
reasonable stabilization in (2+1)D \cite{ber}. A much
better
stabilization results if the Kerr  coefficient is a layered medium
is allowed to vary between successive SDF and SF type
nonlinearities, i.e., between positive and negative values
\cite{mal}. However,  it has been shown that such
a modulation of the  nonlinearity coefficient   in a Kerr medium should
fail to achieve stabilization of a  
light bullet \cite{abdul} or a general three-dimensional soliton
\cite{new}.

As the stabilization of a 
light bullet is of utmost interest, we
revisit this problem and find, to great surprise, that a 
spatiotemporal  light
bullet  can be stabilized in a layered Kerr medium with
sign-alternating nonlinearity   along the
propagation
 direction.

Although, the present work is of interest from a theoretical point of
view, it also has phenomenological or experimental consequences. Recently,
it has been emphasized \cite{liu} that  large negative values of the Kerr
coefficient
can be created by using the cascading mechanism with a large
phase-mismatch parameter. It has also been suggested \cite{xxx} that
a layered medium
with alternating sign of nonlinearity can be created with the technique of
mesoscopic self-organization. Hence a stabilized light bullet can be
experimentally realized in the future.

To stabilize a soliton in a SF homogeneous bulk  Kerr medium,  
the repulsive
kinetic pressure due to the Laplacian operator   in space and time in the
NLS 
has to balance  the attraction due to nonlinearity. For a light
bullet  of
size $L$, kinetic pressure  is proportional to $L^{-2}$ whereas attraction
is
proportional to $L^{-D}$ in $(D+1)$ D. The effective potential,
which
is a sum of these two terms, has a confining minimum only for $D=1$
leading to a stable soliton \cite{ueda}. Using a variational
method 
we find that  a layered
Kerr medium with sign-changing nonlinearity in  (3+1)D 
can lead to an effective potential with a  minimum 
which can
stabilize the solitons.

In Sec. II we present a variational study of the problem and in Sec. III
we present a complete numerical study. Finally, in Sec. IV we give the
concluding remarks.

\section{Variational Calculation}

For anomalous dispersion, the NLS can be written as \cite{1}
\begin{eqnarray}\label{d1} 
 \biggr[ i\frac{\partial}
{\partial z} +\frac{1}{2}\nabla_r^2    +\gamma(z) 
|{u({\bf r},z)}|^2
 \biggr]
u({\bf  r},z)=0,
\end{eqnarray}
where in (3+1)D the three dimensional vector ${\bf r}$
has space components $x$ and $y$ and time component $t$, and $z$ is the
direction of propagation. The Laplacian operator $\nabla_r ^2$ acts on the
variables $x$, $y$, and $t$. In (2+1)D in Eq. (\ref{d1}) the  
vector  ${\bf r}$  could have  components $x$ and
 $y$ or $x$ and $t$, while $z$ continues as  the
direction of propagation.
The nonlinearity coefficient $\gamma(z)$ in a layered Kerr medium is 
 piecewise
continuous and can have  successive positive (SF) and negative
(SDF) values $\gamma_+$ and
$\gamma_-$ in layers of width $d$. 
  The normalization condition  is
$ \int d{\bf r} |u({\bf  r},z)|^2 = P$, 
 where $P$ is the power of the
optical beam \cite{st,mal}.

For a 
spherically symmetric 
soliton in (3+1)D,  $u({\bf r},
z)=
U(r,z) $. Then the radial part of the NLS (\ref{d1}) becomes \cite{1}
\begin{eqnarray}\label{d4}
 \biggr[i\frac{\partial
}{\partial z} +\frac{1}{2}\frac{\partial^2}{\partial
r^2}  +\frac{1}{r}\frac{\partial}{\partial r}           
+ \gamma(z) \left|
{U({r},z)}\right|^2 \biggr] U({r},z)=0.
 \end{eqnarray}
In the following we consider  variational and numerical solutions of
Eq.  (\ref{d4}).

First we consider the variational approach with the following trial 
Gaussian wave function 
for the solution of Eq.   (\ref{d4})  
\cite{ueda,abdul}
\begin{equation}\label{twf}
U(r,z)=N(z)   
\exp\left[-\frac{r^2}{2R^2(z)}
+\frac{i}{2}{ b(z)}r^2+i\alpha(z) 
\right],
\end{equation}
where $N(z)$, $R(z)$, $b(z)$, and $\alpha(z)$ are the soliton's amplitude,
width,
chirp, and
phase, respectively. In
Eq. (\ref{twf}) in
(3+1)D $N(z)={P^{1/2}}/{[\pi^{3/4}R^{3/2}(z)]}$.
 The trial  function   (\ref{twf})  satisfies (a)
the normalization condition  \cite{st,mal}
as well as the boundary conditions
(b) $U(r,z)  \to$ constant  as $r \to 0$ and (c) $|U(r,z)|$ decays
exponentially
as $r \to \infty$ \cite{mal}.
 
The Lagrangian density for
generating Eq.  (\ref{d4})   is \cite{abdul}
\begin{equation}
{\cal L}(U)=\frac{i}{2}\left(\frac{\partial U}{\partial
z}U^*
- \frac{\partial  U^*}{\partial z} U 
\right)-\frac{1}{2}\left|\frac{\partial
 U}{\partial r} \right|^2   
+\frac{\gamma | U|^4}{2} .
\end{equation} 
 The trial  function (\ref{twf}) is
substituted in the Lagrangian density and the 
effective Lagrangian is calculated by
integrating the Lagrangian density: $L_{\mbox{eff}}= \int {\cal L}
(U)
d \bf r.$ The Euler-Lagrange equations for $R(z)$, $b(z)$, and
$\alpha(z)$ are then obtained from
the effective Lagrangian in standard fashion \cite{mal,abdul,ueda}. 
Eliminating $\alpha(z)$,
the equations for $b(z)$ and $R(z)$ in (3+1)D can be written as
\begin{eqnarray} \label{el1}
\frac{d R(z)}{d z} &= &R(z) b (z),\\
\frac{d b(z)}{d z} &= &\frac{1}{R^4(z)} - b^2(z) - \frac{\gamma(z) P}
{2\sqrt{2 \pi^3}}\frac{1}{R^5(z)}. \label{el2} 
\end{eqnarray} 
From Eqs. (\ref{el1}) and (\ref{el2}) we get the following 
second-order differential equation 
 for the evolution of the width
\begin{equation}\label{el3}
\frac{d^2 R(z)}{dz^2}=\frac{1}{R^3(z)}-\frac{\gamma(z) P}{2\sqrt{2
\pi^3}R^4(z)}.
\end{equation}

Here we re-visit the
stability
condition of light bullets of Eq. (\ref{el3}) for $\gamma(z)=\gamma_0+
\gamma_1(z)$, where $\gamma_0=g_0$ is a positive  constant of SF type and
$\gamma_1(z)$ is a
rapidly varying part with zero mean value. We take
$\gamma_1(z)=g_1\sin(\omega z)$, as this is a form that we can integrate
easily.  
 We break
$R(z)$ into a slowly varying part $A(z)$ and a rapidly varying part $B(z)$
by $R(z)=A(z)+B(z)$. Substituting this into Eq. (\ref{el3}) and retaining
terms of the order of $\omega^{-2}$ in $B(z)$ we obtain the following
equations of motion for $B(z)$ and $A(z)$:
\begin{eqnarray}
\frac{d^2 B(z)}{dz^2}&=& -\frac{g_1P\sin (\omega z)}
{2\sqrt{2\pi^3}A^4
(z)}, \end{eqnarray}
\begin{equation}
\frac{d^2 A(z)}{dz^2}=
\frac{1}{A^3(z)}-\frac{g_0 P}{2\sqrt{2\pi^3}A^4
(z)}
+\frac{2 g_1 P\langle{ B(z)\sin(\omega z)}\rangle 
}{\sqrt{2\pi^3}A^5(z)},
\end{equation}
where $\langle\quad \rangle$ denotes time average over rapid
oscillation. Using the solution $B(z)=g_1P\sin (\omega
z)/[2\sqrt{2\pi^3}\omega^2A^4(z)]$, the equation of motion for $A(z)$
becomes 
\begin{eqnarray}\label{ey}
\frac{d^2 A(z)}{dz^2}&=&
\frac{1}{A^3}-\frac{g_0 P  }{2\sqrt{2\pi^3}A^4 }
+\frac{g_1^2 P ^2}{4\pi^3\omega^2 A^9},\\
&=& -\frac{\partial }{\partial A}\left[
\frac{1}{2A^2}-\frac{g_0 P}{6\sqrt{2\pi^3
}A^3}+\frac{g_1^2 P ^2}{32\pi^3\omega^2
A^8} \right].\label{ex}
\end{eqnarray}
The quantity in the square bracket in Eq. (\ref{ex}) is the effective
potential $U(A)$ of the equation of motion
\begin{equation}\label{eff}
U(A)=\frac{1}{2A^2}-\frac{g_0 P}{6\sqrt{2\pi^3
}A^3}+\frac{g_1^2 P ^2}{32\pi^3\omega^2
A^8}.
\end{equation}
 Stabilization is  possible when 
there is a minimum in this effective potential \cite{ueda}. Unfortunately, this
condition
does not lead to a simple analytical solution. However, straightforward
numerical study reveals that this effective potential has a minimum 
for a
positive $g_0   P  $  corresponding to attraction (SF
nonlinearity)
with  
$g_0  P$ above a critical value. For a numerical calculation the quantity 
$gP$ is taken to be of the form $gP=g_0P+g_1P\sin(\omega z)\equiv
g_0P[1+4\sin(10\pi z)]$, so that $g_1=4g_0$ and $\omega = 10\pi$. The
numerical values for $g_1$ and $\omega$ are taken as examples, otherwise
they do not have great consequence on the result so long as $\omega$ is
large corresponding to rapid oscillation. In Fig. 1 (a) we plot the
effective
potential $U(A)$ vs. $A$ for $g_0P=30,100,200$, and 500 for $\omega
= 10\pi$. 
For $g_0P=30$
there is no minimum in $U(A)$, whereas a minimum has appeared for 
 $g_0P=100$ which becomes deeper for  $g_0P=200$ and 500. A careful
examination reveals that the threshold for the minimum in the present case 
is given by  $g_0P\approx 40$. Hence in the present case stabilization is
not possible for $g_0P=30$, and it is possible for $g_0P>40$.        
There is no upper limit for  $g_0 P$ and
stabilization seems
possible for an arbitrarily large   $g_0 P$. As   $g_0 P$ increases the
depth of the effective potential in Fig. 1 (a) increases and consequently,
it
is easier to stabilize a soliton.  As the first
and the third terms on the right hand side (rhs)  of Eq. (\ref{ey}) are
positive, no
stabilization is possible for a negative $g_0P$ corresponding to repulsion
(SDF). 

In order to see the effect of the frequency $\omega$ on
stabilization, we plot in Fig. 1 (b) the same effective potentials of
Fig. 1 (a) for $\omega =30\pi$. With the increase of  $\omega$ the
effective potentials have become deeper and hence the stabilization
easier. For $g_0P=30$ we have a minimum in Fig. 1 (b) whereas there is no
minimum in Fig. 1 (a). With the increase of $\omega$ the threshold value
of $g_0P$ for obtaining a minimum has reduced.

\begin{figure}%[!ht]
 
\begin{center}
\includegraphics[width=1.00\linewidth]{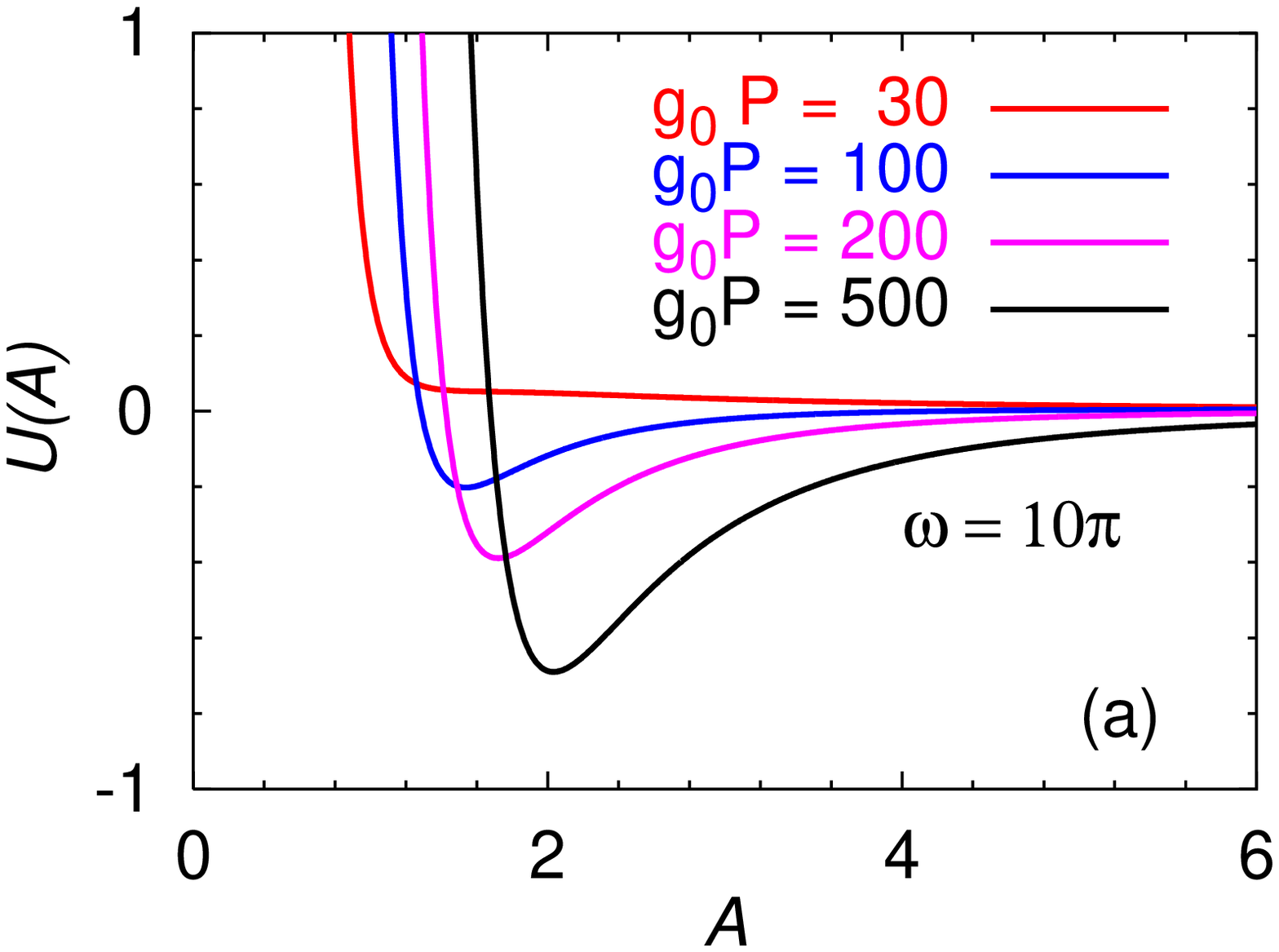}
\includegraphics[width=1.00\linewidth]{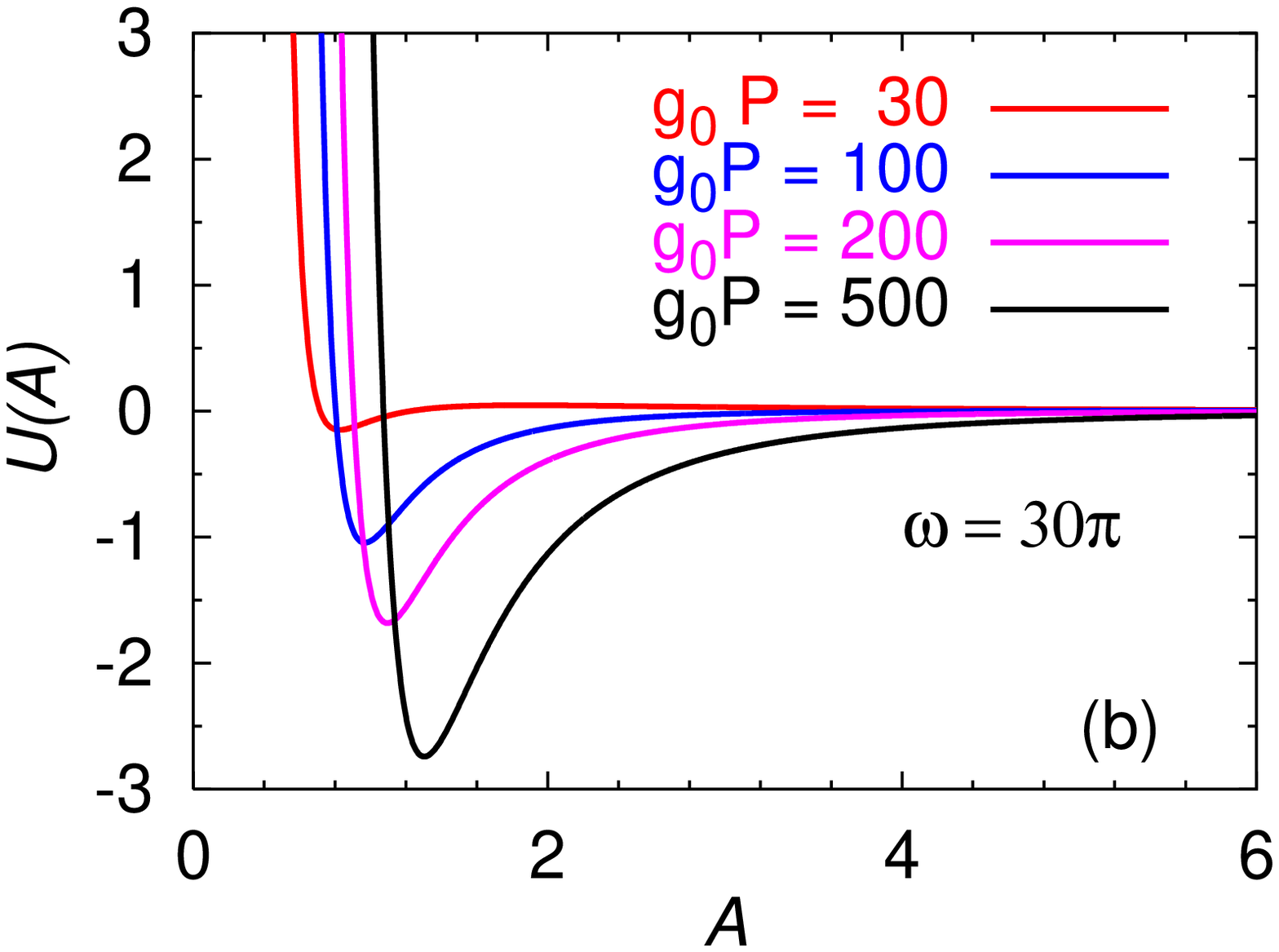}
\end{center}

\caption{The effective potential $U(A)$  of Eq. (12) vs.   $A$ in
arbitrary units for 
$g_0P= 30,100,200,500$ (from shallower to deeper minimum), for
(a) $\omega=10\pi$ and (b) $\omega=30\pi$.  }
\end{figure}

The sinusoidal
variations of
Kerr
nonlinearity as considered above for variational study only simplifies the 
algebra and is by no means necessary for stabilization of solitons. 
In the following numerical study we establish that a rapid oscillation of
the nonlinearity coefficient between positive and negative values  also
stabilizes the soliton in  (3+1)D.

\section{numerical calculation}

We solve Eq.  (\ref{d4}) numerically using  the
split-step time-iteration method employing the Crank-Nicholson
discretization
scheme  \cite{11}. 
The time iteration is started with the
known solution of some auxiliary equation
with zero nonlinearity. 
The auxiliary equations with known Gaussian
solution
are obtained by adding
a harmonic oscillator potential $r^2$ to Eq.  
(\ref{d4}).
Then in the course of time iteration the power $P$ and a  positive 
constant SF Kerr nonlinearity 
$\gamma(z)=g_0=1$
is 
switched on  slowly  and the harmonic trap is also switched off 
slowly.  If the nonlinearity is increased rapidly the system
collapses. The tendency to
collapse or expand   must be avoided to obtain a stabilized
soliton.

After switching off the
harmonic trap 
in Eq.
(\ref{d4}) and after slowly introducing the final power $P$ and the
constant
nonlinearity
$\gamma(z)=g_0=1$,
an oscillating Kerr nonlinearity $\gamma_1(z)$ oscillating between 
$\pm g_1$ $(|g_1|>1)$ like a step function  as $z$ changes by $d$   is
introduced on top of the
constant
nonlinearity. The overall Kerr nonlinearity now has
successive positive and negative values $\gamma_+= 1+g_1$ and  
$\gamma_-= 1-g_1$ of equidistant layers of width $d$ in $z$ direction.
A stabilization of the final solution could be obtained for a suitably
chosen  $\gamma_\pm$ and a small  $d$. 
If the SF
power after switching off the harmonic trap is large compared to  the
spatiotemporal 
size
of the beam
the system becomes highly attractive in the final stage and it eventually
collapses. If the final power  after switching off the harmonic trap 
is small for its size
the system becomes weakly  attractive in the final stage  and it expands. 
The final nonlinearity has to have an appropriate
intermediate value, decided by trial,  
for final stabilization. 
The stabilization could be obtained
for a large range of values of $\gamma_\pm$ and $d$. After some
experimentation with  Eq. 
(\ref{d4}) we opted for the choice $\gamma_+=5$,  $\gamma_-=-3$, and
$d=0.1$ in all our
calculations. 

In the present scheme of stabilization the nonlinearity rapidly fluctuates
between appropriate positive (SF) and negative (SDF) values in the
propagation direction. For positive nonlinearity, the system tends to
collapse, whereas in the SDF regime it tends to expand to infinity. If the
nonlinerities are appropriate, the collapse in the first interval is
exactly compensated for by the expansion in the next interval and a
stabilization of the system is obtained. Obviously, the system will be
more stable when the intervals are small so that  the fluctuation  of the
system around a stable mean position is small. Consequently, the system
remains virtually static and the very small oscillations arising from
collapse and expansion remain unperceptable.

Although, for the sake of convenience we applied a harmonic trap in the
beginning of our simulation, which is removed later with the increase of
nonlinearity, this restriction is by no means necessary for stabilizing a
 soliton. Saito {\it et al.} \cite{ueda} used a qualitatively
similar,
but quantitatively
different, procedure 
for stabilization in the context of Bose-Einstein condensation in two
dimensions. The procedure of  Saito {\it et  al.} could also be applied
sucessfully in the present context.

\begin{figure}%[!ht]
 
\begin{center}
\includegraphics[width=1.00\linewidth]{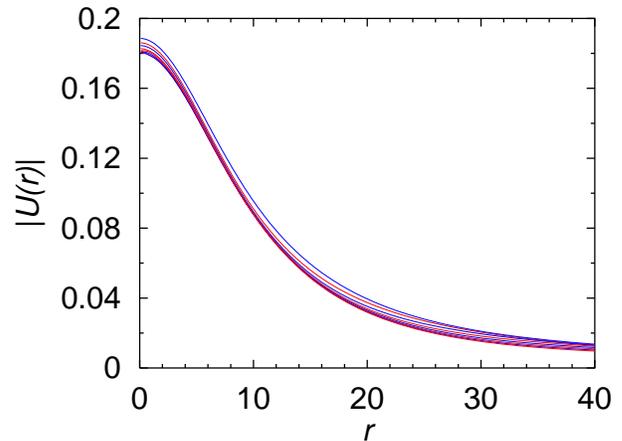}
\end{center}
 
\caption{ Wave function $|U(r)|$ of Eq. (2)  of stabilized light bullet of
power
$ P = 682.6$
and $\langle\gamma(z)\rangle=g_0=1$ and $\gamma_1(z)$ oscillating between 
$\gamma_+=5$ and $\gamma_-=-3$ in successive layers of 
width   $d=0.1$   in the $z$ direction.
The wave functions are shown 
at positions
$z=0,50,100,150, 200,250,300,350 $ and 400.}
\end{figure}

Now  we turn to a numerical
investigation of Eq. (\ref{d4}).
The results are shown in Fig. 2  for the (3+1)
dimensional soliton, where we plot the
radial part of the wave function for different $z$ for a power $P=682.6$.
A finetuning of the power was needed for the stabilization reported in
Fig. 2. 

In Fig. 2   
the narrow spread of the wave
function over the large
interval of $z$ shows the quality of stabilization.
The results at intermediate $z$ lie in the region covered by the 
plots.
The plot of the full wave function at different $z$ on the same 
graph clearly
shows the degree
of stabilization achieved. The stabilization seems to be perfect 
and can easily 
be continued for longer intervals of  $z$ by increasing the power. In
Ref. \cite{mal} layers of
width $d=0.001$ was employed for stabilization in (2+1)D. 
The present stabilization is obtained with a much larger
width $d=0.1$, which make the present proposal more attractive from a
phenomenological point of view.  The stabilization can only be 
obtained for beams with power larger than a critical value. 
Numerically, we found 
it was easier to obtain stabilization of beams with power much larger than
the critical value. 
In (3+1)D
good stabilization could be obtained for much larger power: the power
employed in Fig. 2 was 682.6, whereas the critical variational power for
stabilization obtained in Fig. 1 is about 40.

Using a variational procedure alone, not quite identical with the present
approach, in the context of Bose-Einstein condensation Abdullaev {\it et
al.} \cite{abdul} also had found that a
stabilization of a soliton could be possible in (3+1)D via a temporal
modulation of the nonlinearity. However, they confirmed after further
analytical and numerical study that such a stabilization does not take
place in (3+1)D. Saito {\it et al.} \cite{ueda} and Towers  {\it et al.}
\cite{mal}, on the other hand, are silent
about the possibility of the stabilization of a soliton in (3+1)D.  We
point
out one possible reason for the negative result obtained by Abdullaev {\it
et al.} \cite{abdul} in (3+1)D. The nonlinearity parameter
$\lambda_0 N=\sqrt 2 \pi^{3/2}\Lambda$ with $\Lambda=1$ 
used
in Ref. \cite{abdul} for stabilizing a soliton in (3+1)D 
is much too small (smaller than the threshold discussed in
Sec. II). Comparing
Eq. (2)  of \cite{abdul} with our Eq. (\ref{d4}) we find that the above
value of nonlinearity 
corresponds  in our
notation to  $g_0P=\lambda_0 N/2= \pi^{3/2}/\sqrt 2\approx 3.9$, whereas
the
present variational  threshold for obtaining a stabilized soliton is 
$g_0P\approx 40$. The very small nonlinearity used in Ref. \cite{abdul} is
most possibly the reason for the negative result obtained there.
More recently, Montesinos {\it et al.} \cite{new} have also confirmed the
conclusion of  Ref. \cite{abdul} that no stable three-dimensional soliton
could be obtained by a variation of Kerr nonlinearity. However, they did
not give details of their study, which led to this conclusion,  for a
comparison.

\section{Conclusion}

In conclusion,  after a variational and numerical study of the NLS we find
that 
it is possible to stabilize 
a spatiotemporal light bullet  in (3+1)D 
by employing a layered Kerr
medium with a sign-changing nonlinearity along the propagation direction.
From a variational calculation we
show that a oscillating Kerr nonlinearity produces a
minimum in the effective potential, thus producing a
potential well in
which the soliton can
be trapped. The present stabilized soliton is a slowly
collapsing Townes soliton [1] with large power. The oscillating Kerr
nonlinearity stops the collapse and enhances the lifetime of
the soliton
greatly.
This is
of interest to investigate if such light  bullets 
could be created
experimentally. 

Apart from optics, such stable three-dimensional solitons can be realized
experimentally in Bose-Einstein condensates (BEC), where a Feshbach
resonance
could be used to generate an oscillating nonlinearity or an oscillating 
effective interatomic interaction via the modulation 
of an external background magnetic field \cite{abdul,ueda}. The
stabilization
of such BEC solitons in two \cite{abdul,ueda,new} and three \cite{unp}
dimensions 
is already under investigation. 

\acknowledgments

I thank  Dr. R. A. Kraenkel 
for informative discussions. 
The work was supported in part by the CNPq 
of Brazil.


\begin{thebibliography}{99}

\bibitem{st} R. Y. Chiao, E. Garmire, and C. H. Townes, Phys. Rev. Lett.
{\bf 13}, 479 (1964). 


\bibitem{0}A.
 Hasegawa  and F. Tappert,  Appl. Phys. Lett. {\bf 23}, 171 (1973),
{\bf 23}, 142 (1973);
V. E. Zakharov   and A. B. Shabat    Sov. Phys. JETP   {\bf 34},
62 (1972), {\bf 37},  823 (1973).

 
\bibitem{1}Y. S. Kivshar and G. P. Agrawal, {\it Optical Solitons - From
Fibers to Photonic Crystals}, (Academic Press, San Diego, 2003). 



\bibitem{2} V. I. Kruglov and R. A. Vlasov, Phys. Lett. {\bf 111A}, 401
(1985).




\bibitem{3}W. J. Firth and D. V. Skryabin, Phys. Rev. Lett. {\bf 79}, 2450
(1997);
D. V. Skryabin and W. J. Firth, Phys. Rev. E {\bf 58}, 3916 (1998);
J. Atai, Y. Chen, and J. M. Soto-Crespo,  Phys. Rev. A {\bf 49}, R3170
(1994).

\bibitem{3a} V. V. Afanasjev,  Phys. Rev. E {\bf 52}, 3153 (1995).


\bibitem{mal}
I. Towers and B. A. Malomed, J. Opt. Soc. Am. B {\bf 19}, 537 (2002). 

\bibitem{ber}L. Berg\'e, V. K. Mezentsev, J. J. Rasmussen,
P. L. Christiansen, and Yu. B. Gaididei, Opt. Lett. {\bf 25}, 1037 (2000).


\bibitem{abdul}F. K. Abdullaev, J. G. Caputo, R. A. Kraenkel, and 
B. A. Malomed, Phys. Rev. A {\bf 67}, 013605 (2003).

\bibitem{new} G. D. Montesinos, V. M. Perez-Garcia, and  P. J. Torres,
Physica D {\bf 191}, 193 (2004).


\bibitem{liu}L. J. Qian, X. Liu, and F. W. Wise, Opt. Lett. {\bf 24}, 166
(1999); X. Liu, L. J. Qian,  and F. W. Wise, {\it ibid.}
 {\bf 24}, 1777 (1999).

\bibitem{xxx} L. Brzozowski and E. H. Sargent, IEEE J. Quantum
Electron. {\bf 36}, 550   (2000).




\bibitem{ueda}H. Saito and M. Ueda, Phys. Rev. Lett. {\bf 90}, 040403
(2003).


\bibitem{11} S. K.
Adhikari and P.  Muruganandam,  {
J. Phys. B} {\bf 35}, 2831 (2002); 
{\bf 36}, 409 (2003);
P.  Muruganandam  and S. K.  Adhikari, {\it ibid.}  
{\bf 36}, 2501 (2003).


\bibitem{unp}S. K. Adhikari, Phys. Rev. A {\bf 69}  063613 (2004).

 
 
 
 
 
 
 
 
\end{thebibliography}
\end{document}